\documentclass[aps, pra, a4paper,twocolumn, showpacs,10pt]{revtex4-1}
\usepackage{bbm,amsmath,amssymb,amsthm,bm, bbold,textcomp,graphicx,epsfig,color,verbatim,enumerate}
\usepackage{upgreek}

\newcommand{\ket}[1]{\left|#1\right\rangle}
\newcommand{\bra}[1]{\left\langle #1\right|}
\newcommand{\bracket}[2]{\left\langle #1|#2\right\rangle}
\newcommand\defn[1]{\textsl{#1}}

\newcommand\ketbra[2]{|#1\rangle\langle#2|}

\newcommand{\one}{\mathbb{1}}

\def\sx{\sigma_x}
\def\sy{\sigma_y}
\def\sz{\sigma_z}
\def\sp{\sigma_+}
\def\sm{\sigma_-}

\makeatletter
\newtheorem*{rep@theorem}{\rep@title}
\newcommand{\newreptheorem}[2]{%
\newenvironment{rep#1}[1]{%
 \def\rep@title{#2 \ref{##1}}%
 \begin{rep@theorem}}%
 {\end{rep@theorem}}}
\makeatother

\newreptheorem{theorem}{Theorem}
\newreptheorem{lemma}{Lemma}

\begin{document}

\date{\today}

\author{Michael Skotiniotis}
\affiliation{Institut f\"ur Theoretische Physik, Universit\"at Innsbruck, Technikerstr. 25, A-6020 Innsbruck, Austria}
\author{Pavel Sekatski}
\affiliation{Institut f\"ur Theoretische Physik, Universit\"at Innsbruck, Technikerstr. 25, A-6020 Innsbruck, Austria}
\author{Wolfgang D\"ur}
\affiliation{Institut f\"ur Theoretische Physik, Universit\"at Innsbruck, Technikerstr. 25, A-6020 Innsbruck, Austria}

\title{Quantum metrology for the Ising Hamiltonian with transverse magnetic field}

\begin{abstract}
We consider quantum metrology for unitary evolutions generated by parameter-dependent 
Hamiltonians.  We focus on the unitary evolutions generated by the 
Ising Hamiltonian that describes the dynamics of a one-dimensional chain of spins with nearest-neighbour 
interactions and in the presence of a global, transverse, magnetic field. We analytically solve the problem and show that the 
precision with which one can estimate 
the magnetic field (interaction strength) given one knows the interaction strength (magnetic field) scales at the 
Heisenberg limit, and can be achieved by a linear superposition of the vacuum and $N$ free fermion states.  In addition, 
we show that GHZ-type states exhibit Heisenberg scaling in precision throughout the entire regime of parameters.  Moreover, we 
numerically observe that the optimal precision using a product input state scales 
at the standard quantum limit.  
\end{abstract}
\pacs{03.67.-a, 06.20-f, 75.10.Pq}
\maketitle

\section{Introduction}
\label{intro}

Quantum metrology is one of the archetypical applications where quantum mechanics demonstrably exhibits a vast 
improvement over the best known classical strategies.  The use of entangled input states of $N$ qubits, such as GHZ 
states~\cite{GHZ:89}, is known to allow for a precise determination of an unknown 
parameter such as the relative phase in a Mach-Zender interferometer~\cite{Holland:93, Hwang:02, Dorner:09},
or the frequency of an atomic transition~\cite{Bollinger:96, Dorner:12} with 
a precision that scales inversely proportional to $N$, the \defn{Heisenberg limit}.  In comparison the best 
classical estimation strategy, which employs separable states, gives a precision scaling inversely 
proportional to $\sqrt{N}$, the \defn{standard quantum limit}~\cite{GLM:04, GLM:06}.  This observation has proven 
very useful in the development of ultra-precise atomic clocks~\cite{Borregaard:13, Kessler:13, Rosenband:13}, 
high resolution imaging~\cite{Low:14}, as well as the detection of gravitational waves~\cite{McKenzie:02,Ligo:11}.

In the absence of any noise or decoherence effects the parameter of interest, $\lambda$, in a metrological scenario is 
imprinted onto the state of $N$ probes via the unitary operator $U=e^{iH(\lambda)t}$, where $H(\lambda)$ is the 
Hamiltonian describing the dynamics of the $N$ probes. For the case of phase and frequency estimation---where the 
parameter to be estimated is $\lambda t$ and $\lambda$ respectively---the parameter is a \defn{multicplicative factor} 
of the Hamiltonian, $H(\lambda)=\lambda H$, and the Hamiltonian is \defn{local}, i.e.,~$H=\sum_{i=1}^N h^{(i)}$ 
where $h^{(i)}$ is the Hamiltonian describing the evolution of the $i^{\mathrm{th}}$ probe.  
Indeed, quantum metrology using local Hamiltonians where the parameter of interest enters only as a multiplicative 
factor have been studied extensively both in the absence as well as in the presence of 
noise~\cite{Huelga:97, Escher:11, Demkowicz:12, Kolodynski:13, Alipour:13, Benatti:14, Knysh:11, Knysh:14, FSKD:14,*SFKD:14}.

In stark contrast quantum metrology with more general Hamiltonians is only now beginning to attract attention.
Some instances of quantum metrology with parameter dependent Hamiltonians concern the estimation of time-varying 
signals~\cite{Tsang:11, Latune:12, Magesan:13}, the estimation of magnetic-field gradients along a spin 
chain~\cite{Lanz:13, NK:14, Zhang:14}, or the estimation of the anisotropy and/or  decoherence along a spin-chain using 
nonequilibrium states~\cite{Marzolino:14}. The general problem of performing quantum metrology using parameter 
dependent Hamiltonians was treated in~\cite{DePasquale:13,Pang:14} where it was shown that for parameter dependent 
local Hamiltonians Heisenberg scaling precision with respect to the number of probing systems is possible. 

In this work we focus on noiseless quantum metrology of $N$ interacting probe systems described by the Ising Hamiltonian
\begin{equation}
H(J,B)=J\sum_{i=1}^{N-1}\sx^{(i)}\sx^{(i+1)} + B\sum_{i=1}^N\sz^{(i)},
\label{Ising}
\end{equation}  
where we are interested in determining the precision with which one can estimate either the strength of the transverse 
magnetic field, $B$, or the coupling interaction, $J$, provided the remaining quantity is known.  The Hamiltonian of 
Eq.~\eqref{Ising} is known to exhibit a phase transition~\cite{Sachdev:07}, which have been discussed previously in 
the context of quantum metrology and were shown to be 
resourceful~\cite{Zanardi:07, Zanardi:08, Gammelmark:13,Macieszczak:14}.
Moreover, as the Ising Hamiltonian is entanglement generating~\cite{Schuch:08}, it has found applications
in ion-trap quantum computing architectures~\cite{Jurcevic:14, Schachenmayer:13}, where either $J$ or $B$ can 
be controlled at will by modifying either the separation of the ions, or the global magnetic field strength.

Here we report the following results regarding precise estimation of parameters of the Ising Hamiltonian:
\begin{enumerate}
\item The ultimate precision with which one can estimate either $J$ or $B$, having complete knowledge of $B$ or $J$,
using $N$ probe systems scales at the Heisenberg limit, and we provide an analytic expression for this achievable 
precision as well as the optimal states that achieve it.  Our result answers the conjecture in~\cite{DePasquale:13} that 
Heisenberg scaling is the ultimate limit also for the case of $N$ interacting probes for the case of nearest-neighbour 
interactions. 
\item  We provide numerical evidence that the ultimate achievable precision strictly outperforms the optimal classical 
strategy, which deploys the $N$ initial probes in a pure product state.  For up to $N=11$ our numerical study shows that the optimal 
product input state yields a precision that scales at the standard quantum limit.
\item  We analytically derive the precision achieved by the GHZ-type states, known to achieve the optimal precision when either $J$ 
or $B$ are equal to zero, and show that these states retain their Heisenberg scaling in precision (up to a constant factor) over the
entire regime of parameters.
 \end{enumerate}

This paper is organised as follows.  In Sec.~\ref{background} we review the basics of quantum metrology as well as 
some important mathematical results regarding unitary operators generated by parameter dependent Hamiltonians.  
In Sec.~\ref{Estimation} we use the Jordan-Wigner transformation to determine the maximal possible precision with 
which one can estimate either $J$ and $B$ using $N$ systems as probes. We compare this to the best possible 
precision that can be achieved by a separable state of $N$ probes which we determine numerically for up to $N=11$ 
qubits.  In Sec.~\ref{GHZprecision} we analytically determine the performance of states that are known to be optimal at the 
extreme cases where $J=0$ and $B=0$.  Finally, Sec.~\ref{Conclusion} contains the conclusions of our investigation as 
well as some open questions for future work.
 
\section{Basics of Quantum Metrology}
\label{background}

In this section we review the main results in quantum metrology and review some important facts pertaining to 
parameter-dependent Hamiltonians in general and to the Ising Hamiltonian (Eq.~\eqref{Ising}) in particular.
Specifically, we concentrate on noiseless quantum metrology and the quantum Fisher information (QFI); the central 
quantity of interest in quantum metrology.  After introducing the QFI we provide a formula for calculating it for the 
case of general parameter-dependent Hamiltonians and then to the specific case of the Ising Hamiltonian.  We note that
the behaviour of the QFI, and in particular its scaling with time and number of probe systems, was investigated for general 
parameter dependent Hamiltonians in~\cite{DePasquale:13, Pang:14}.

A standard protocol in noiseless quantum metrology can be formulated as follows: $N$ probes are prepared in a 
suitable state and undergo an evolution for some time, $t$, described by the \defn{unitary} operator
$U(\bm\lambda, t)=e^{-i t H(\bm\lambda)}$, where $H(\bm\lambda)$ is the Hamiltonian describing the dynamics of the 
$N$ probes and explicitly depends on the vector of parameters $\bm\lambda\equiv(\lambda_1,\ldots,\lambda_M)$.  
Finally the $N$ probes are measured and an estimate of $\bm{\hat\lambda}$ is obtained from the measurement 
statistics of $\nu$ repetitions of the above procedure.  In what follows we shall assume that all other parameters 
except $\lambda_i$ are known, and shall be concerned with estimating $\lambda_i$ as precisely as possible.   A lower 
bound on the error, $\delta\lambda_i=\lambda_i-\hat\lambda_i$, for any 
unbiased estimator $\hat\lambda_i$ is given by the quantum Cram\'{e}r-Rao bound~\cite{BC:94}
\begin{equation}
\delta \lambda_i \geq \frac{1}{\sqrt{\nu {\cal F}(\rho_{\bm{\lambda},t})}},
\label{QCR}
\end{equation}
where ${\cal F}(\rho_{\bm{\lambda},t})$ is the \defn{quantum Fisher information} (QFI) of the state
$\rho_{\bm{\lambda},t}$ describing the $N$ probes after the unitary dynamics have acted.  In the most general case 
the QFI can be computed as~\cite{BC:94}
\begin{equation}
\mathcal{F}(\rho_{\bm{\lambda},t})=\mathrm{Tr}\left\lbrace L_{\lambda_i}\rho_{\bm{\lambda}, t}L_{\lambda_i}\right\rbrace,
\label{QFI}
\end{equation}
with
\begin{equation}
L_{\lambda_i}=2\sum_{\alpha,\beta}\frac{\bra{\psi_\alpha}\partial_{\lambda_i}{\rho}_{\bm{\lambda}, t}\ket{\psi_\beta}}{\alpha+\beta}\ket{\psi_\alpha}\bra{\psi_\beta}
\label{SLD}
\end{equation}
the \defn{symmetric logarithmic derivative}, where $\alpha\, (\beta)$ are the eigenvalues of $\rho_{\bm{\lambda}, t}$, 
$\ket{\psi_\alpha}, (\ket{\psi_\beta})$ the corresponding eigenvectors, and the sum in Eq.~\eqref{SLD} is over 
all $\alpha,\, \beta$ satisfying $\alpha+\beta\neq0$.  Here and in what follows 
$\partial_x\equiv \frac{\partial}{\partial x}$.  We now review the case where the parameter of interest enters as a multiplicative factor of the Hamiltonian

\subsection{Parameter independent Hamiltonians}
In the case of noiseless metrology, an easier expression for computing the QFI exists if one initializes the $N$ probes 
in a pure state $\rho=\ketbra{\psi}{\psi}$.  In this case the QFI can be shown to be
\begin{equation}
\mathcal{F}\left(\ket{\psi_{\bm{\lambda},t}}\right)=4\left(\bracket{\partial_{\lambda_i}{\psi}_{\bm{\lambda}, t}}{\partial_{\lambda_i}{\psi}_{\bm{\lambda}, t}}-\left|\bracket{\partial_{\lambda_i}{\psi}_{\bm{\lambda}, t}}{\psi_{\bm{\lambda}, t}}\right|^2\right),
\label{QFIpure}
\end{equation} 
where $\ket{\psi_{\bm{\lambda}, t}}=U(\bm{\lambda}, t)\ket{\psi}$. If $H(\bm{\lambda})=\lambda_i H$ 
then a bit of algebra yields $\mathcal{F}\left(\ket{\psi_{\bm{\lambda},t}}\right)=4 t^2 \Delta^2 H$, where 
$\Delta^2 H\equiv \bra{\psi} H^2\ket{\psi}-\left|\bra{\psi} H\ket{\psi}\right|^2$ is the variance of $H$ with respect to the 
state $\ket{\psi}$.  If, in addition, $H=\sum_{i=1}^N h^{(i)}$ where $h^{(i)}=h^{(j)}=h,\, \forall i,j$ is the Hamiltonian 
acting on the $i^{\mathrm{th}}$ probe system, then it can be shown that when 
$\ket{\psi}=(\frac{\ket{\alpha_{\min}}+\ket{\alpha_{\max}}}{\sqrt{2}})^{\otimes N}$, where 
$\ket{\alpha_{\min\,(\max)}}$ are the eigenstates corresponding to the minimum and maximum eigenvalues of $h$,  
then $\mathcal{F}\left(\ket{\psi_{\bm{\lambda},t}}\right)=t^2 N (\alpha_{\max} - \alpha_{\min})^2$ and give the 
\defn{standard quantum limit} in estimation precision~\cite{GLM:06}.  On the other hand if the probes are prepared in 
the Greenberger-Horne-Zeilinger (GHZ) state
\begin{equation}
\ket{\psi}=\frac{1}{\sqrt{2}}(\underbrace{\ket{\alpha_{\min},\,\ldots\,,\alpha_{\min}}}_{N\, \mathrm{times}}+\underbrace{\ket{\alpha_{\max},\,\ldots\,,\alpha_{\max}}}_{N\, \mathrm{times}}),
\label{ghz}
\end{equation}
then $\mathcal{F}\left(\ket{\psi_{\bm{\lambda},t}}\right)=t^2 N^2 (\alpha_{\max} - \alpha_{\min})^2$, the Heisenberg 
scaling in estimation precision~\cite{GLM:06}. In general, Heisenberg scaling in precision implies an improvement in scaling with regards to the number of probe systems, over the optimally achievable precision which uses the same number 
of probe systems initialized in a separable state.  Hence, in the case where the parameter to be estimated is a 
multiplicative factor of a local Hamiltonian, the use of highly entangled states leads to a quadratic improvement in 
scaling precision.  Note that besides the GHZ states there exists a large class of pretty good states that scale at the 
Heisenberg limit up to a multiplicative factor~\cite{SFKD:14}. 

\subsection{Parameter dependent Hamiltonians}

We now  use Eq.~\eqref{QFIpure} to compute the QFI for Hamiltonians of the form
\begin{equation}
H(\bm{\lambda})=\lambda_1 H_1+\lambda_2 H_2
\label{general H}
\end{equation}
where $[H_1, H_2]\neq 0$.  We make no assumptions on the structure of $H_1, \, H_2$; in particular we do not 
assume that they are local Hamiltonians. Notice that the Ising Hamiltonian of Eq.~\eqref{Ising} is a special case of 
Eq.~\eqref{general H}, whose QFI was studied in detail in~\cite{DePasquale:13} with 
$\lambda_1 H_1= J\sum_{i=1}^N\sx^{(i)}\sx^{(i+1)}$ and $\lambda_2 H_2=B\sum_{i=1}^N\sz^{(i)}$.  
From Eq.~\eqref{QFIpure} we need to compute 
$\ket{\partial_{\lambda_i}\psi_{\bm{\lambda}, t}}=\partial_{\lambda_i} U(\bm{\lambda},t)\ket{\psi}$.  
As $[H_1, H_2]\neq0$ and using 
\begin{equation}
\frac{\partial e^{-i\,t(\lambda_1 H_1+\lambda_2 H_2)}}{\partial_{\lambda_i}}=\lim_{N\to\infty}\frac{\partial\left(\one-\frac{i\, t}{N}(\lambda_1H_1+\lambda_2 H_2)\right)^N}{\partial_{\lambda_i}},
\label{lieformula}
\end{equation} 
Eq.~\eqref{QFIpure} reads  
\begin{align}\nonumber
\mathcal{F}\left(\ket{\psi_{\bm{\lambda},t}}\right)&=4\left(\bra{\psi}U^\dagger(\bm{\lambda},t)\mathcal{O}_i^2(\bm{\lambda}, t)U(\bm{\lambda},t)\ket{\psi}\right.\\ \nonumber
&\left.-\left|\bra{\psi}U^\dagger(\bm{\lambda},t)\mathcal{O}_i(\bm{\lambda}, t)U(\bm{\lambda},t)\ket{\psi}\right|^2\right)\\
&=4\Delta^2\mathcal{O}_i(\bm{\lambda},t),
\label{QFIising}
\end{align}
where 
\begin{equation}
\mathcal{O}_i(\bm{\lambda}, t)\equiv\int_0^t\, \mathrm{d}s\, U(\bm{\lambda},s) H_i U^\dagger(\bm{\lambda},s)
\label{operator}
\end{equation}
and the variance of $\mathcal{O}_i$ is computed with respect to the evolved state $\ket{\psi_{\bm{\lambda},t}}$.

Thus, the QFI is maximised by the states that are linear 
superpositions of the eigenstates corresponding to the minimum and maximum eigenvalues of 
$\mathcal{O}_i(\bm{\lambda},t)$.  An interesting question is whether the presence of $\lambda_1H_1$ ($\lambda_2H_2$) can help 
boost the precision of estimation of parameters for $\lambda_2$ ($\lambda_1$) respectively.  
It was shown in~\cite{DePasquale:13} that this is not the case.  Specifically, if one has control over either 
$\lambda_1H_1$ or $\lambda_2H_2$ and wishes to estimate 
$\lambda_2$ or $\lambda_1$ respectively, then the optimal strategy is to set the dynamics under our control to 
zero.

In the next section we use the expressions in Eq.~\eqref{QFIising} to determine the optimal QFI for 
either the magnetic field $B$ or interaction strength $J$ of the Ising Hamiltonian using separable and entangled states 
respectively.
 
\section{Estimation of magnetic field and interaction strength}
\label{Estimation}

We are interested in determining the optimal precision in estimating either the magnetic field strength, $B$, or 
interaction strength, $J$, of the Ising Hamiltonian (Eq.~\eqref{Ising}).  In particular, we will show that the optimal 
precision in estimating either $B$ or $J$ scales at the Heisenberg limit, up to a constant factor which depends only 
on the ratio of $J$ and $B$, and is achievable by states that are linear 
superpositions of the vacuum and fully occupied states of free fermions of a suitable type.  
Furthermore, we numerically determine the best achievable precision using a separable state for up to $11$ qubits 
and show that the optimal precision scales, to within best fit errors, linearly with $N$.  Hence, our results provide 
strong evidence that the entanglement generated by the Ising Hamiltonian when acting on an initially pure separable 
state is not enough to boost the precision in estimation from the SQL to the Heisenberg limit. 

We begin by first determining the optimal precision in estimation of either $B$, or $J$, and the corresponding optimal 
states.  To do so we note that via the use of the Jordan-Wigner transformation~\cite{Jordan:28} the Ising Hamiltonian 
of Eq.~\eqref{Ising} can be expressed as a quadratic Hamiltonian in fermionic creation and annihilation operators 
which can be suitably diagonalized.  Specifically the mapping
\begin{align}\nonumber
a_j&\equiv\left(\bigotimes_{k=1}^{j-1}\sz^{(k)}\right)\otimes \sm^{j}\\
a_j^{\dagger}&\equiv\left(\bigotimes_{k=1}^{j-1}\sz^{(k)}\right)\otimes \sp^{j}
\label{JW1}
\end{align}
and its inverse
\begin{align}\nonumber
\sm^{(j)}\equiv\exp\left(i\pi\sum_{k=0}^{j-1}a_k^{\dagger}a_k\right)a_j\\
\sp^{(j)}\equiv\exp\left(i\pi\sum_{k=0}^{j-1}a_k^{\dagger}a_k\right)a_j^{\dagger},
\label{JW2}
\end{align}
where $\{\sx, \,\sy,\, \sz\}$ are the Pauli matrices~\footnote{We note that since we are dealing with spin-$1/2$ systems 
the Pauli matrices are defined as $\sx=\frac{1}{2}\left(\begin{smallmatrix}0&1\\1&0\end{smallmatrix}\right)$ and 
likewise for $\sy,\,\sz$.}, with $\sigma^{(j)}_\pm\equiv\sx^{(j)}\pm i\sy^{(j)}$, and
$a_j, \,a_j^{\dagger}$ are the fermionic annihilation and creation operators for mode $j$ 
respectively and satisfy the anti-commutation relations 
$\{a_j^{\dagger}\,a_k^\dagger\}=\{a_j\,a_k\}=0$, $\{a_j,\, a_k^\dagger\}=\delta_{jk}\one$.   
Substituting Eq.~\eqref{JW2} into Eq.~\eqref{Ising} yields 
\begin{equation}
H(J,B)=J\sum_{j=1}^N(a_j^\dagger-a_j)(a_{j+1}^\dagger+a_{j+1})+ 2B\sum_{j=1}^Na_j^\dagger a_j.
\label{Ising_JW}
\end{equation}

Any quadratic Hamiltonian in the fermionic operators can be brought to the diagonal form 
\begin{equation}
\tilde{H}(J, B)=2\sum_{k=0}^{N-1} \sqrt{\alpha_k^2+\beta_k^2}\quad c_k^\dagger c_k,
\label{Isingfreefermion}
\end{equation} 
where $\alpha_k=J\cos\left(\frac{2\pi k}{N}\right)+B$, $\beta_k=J\sin\left(\frac{2\pi k}{N}\right)$, and $c^{\dagger}_k$, 
$c_k$ are the creation and annihilation operators of free fermions in mode $k$, with $c_N=c_0$.  The eigenstates 
of $\tilde{H}(J,B)$ are \defn{fermionic Fock states}, $\ket{\bm k}$, where $\bm k$ is an $N$-bit binary string indicating 
which modes are occupied by fermions.  Without loss of generality we may set the 
\defn{vacuum state}, $\ket{\bm 0}$ to have energy $\sqrt{\alpha_0^2+\beta_0^2}=0$.  The maximally occupied 
fermionic Fock state, $\ket{\bm 1}$ has energy equal to $\sum_{k=0}^{N-1}\sqrt{\alpha_k^2+\beta_k^2}$.

As Eq.~\eqref{Isingfreefermion} is of great importance in the remainder of this work, we now discuss  
the steps required for obtaining it.  Starting from the quadratic Hamiltonian of Eq.~\eqref{Ising_JW} one first 
performs the \defn{Fourier transformation}
\begin{equation}
b_j=\frac{1}{\sqrt{N}}\sum_{k=0}^{N-1}\,e^{-i\frac{2\pi j k}{N}}a_k,
\label{FT}
\end{equation}
of the mode operators $a_k$.  After substituting Eq.~\eqref{FT} into Eq.~\eqref{Ising_JW} the Hamiltonian can be 
written in \defn{matrix form} as 
\begin{equation}
\tilde{H}(J,B)=\sum_{k=0}^{N-1}\left(b_k^\dagger, \, b_{N-k}\right)\left(\begin{matrix} \alpha_k & i\beta_k\\-i\beta_k & -\alpha_k\end{matrix}\right) \left(\begin{matrix} b_k\\ b^\dagger_{N-k}\end{matrix}\right).
\label{Hamiltonianmatrixform}
\end{equation}
This block-diagonal structure of the Hamiltonian in terms of the mode operators $b_k$ and $b_{N-k}$ makes it 
particularly useful for 
calculating the operator $\mathcal{O}$ of Eq.~\eqref{operator}. Finally, from Eq.~\eqref{Hamiltonianmatrixform} one 
can go to the diagonalized Hamiltonian of Eq.~\eqref{Isingfreefermion} by performing a suitable Bogoliubov transformation on 
the two-dimensional block of mode operators $b_k,\, b_{N-k}$
\begin{align}\nonumber
c_k=\cos \theta_k\, b_k -e^{i\phi_k}\sin\theta_k\,b^\dagger_{N-k}\\
c_{N-k}=e^{i\phi_k}\sin\theta_k\,b^\dagger_k+\cos \theta_k\, b_{N-k}
\label{BT}
\end{align}
with $\theta_k=\frac{1}{2}\tan^{-1}\left(\frac{J\sin\left(\frac{2\pi k}{N}\right)}{J\cos\left(\frac{2\pi k}{N}\right)+B}\right)$ 
and $\phi_j=\frac{\pi}{2}, \forall j$.

We now determine the optimal achievable precision in estimating $J$ given that $B$ is known.

\subsection{Estimating the Interaction strength}
\label{estimatingJ}
We now proceed to estimate the interaction strength $J$ for the Ising Hamiltonian.  The QFI is given by 
Eq.~\eqref{QFIising} with
\begin{equation}
\mathcal{O}_J(J,B, t)=\int_0^t U(J,B,s) H_1 U^\dagger(J,B,s)\,\mathrm{d}s,
\label{Interaction}
\end{equation}
where $H_1=\sum_{i=1}^{N}\sx^{(i)}\sx^{(i+1)}$.  Writing the latter in terms of the fermionic operators $b_k$ one 
obtains
\begin{equation}
H_1=\sum_{k=0}^{N-1}\left(\begin{matrix}b_k^\dagger & b_{N-k}\end{matrix}\right)\underbrace{\left(\begin{matrix} \cos\left(\frac{2\pi k}{N}\right) & i\sin\left(\frac{2\pi k}{N}\right)\\ -i\sin\left(\frac{2\pi k}{N}\right)  & -\cos\left(\frac{2\pi k}{N}\right)\end{matrix}\right)}_{M_k} \left(\begin{matrix} b_k\\ b^\dagger_{N-k}\end{matrix}\right).
\label{h1intermsofb}
\end{equation}

In order to calculate the operator $\mathcal{O}_{J}(J,B,t)$ of Eq.~\eqref{Interaction} we need to determine the action 
of $H(J,B)$ on the fermionic operators $b_k$, i.e.,~we need to determine $b_k(s)=U(J,B, s) b_k(0) U^\dagger(J,B,s)$.  
This is simply the Heisenberg equation of motion for the mode operators $b_k$.  Due to the 
block-diagonal structure of $\tilde{H}(J,B)$, when written in terms of fermionic operators $b_k$, the solution to the 
Heisenberg equation of motion can be seen to be
\begin{equation}
\left(\begin{matrix} b_k(s) \\ b_{N-k}^\dagger(s)\end{matrix}\right) =
\underbrace{\exp\left(-i\,s\left(\begin{matrix} \alpha_k & i\beta_k\\-i\beta_k & -\alpha_k\end{matrix}\right)\right)}_{R_k(s)}
\left(\begin{matrix} b_k(0) \\ b_{N-k}^\dagger(0)\end{matrix}\right).
\label{actionHonb}
\end{equation} 
Henceforth, we drop the explicit time dependence of the mode operators $b_k$ for convenience.

Substituting Eq.~\eqref{actionHonb} into Eq.~\eqref{Interaction} yields
\begin{equation}
\mathcal{O}_J(J,B,t)=\sum_{k=0}^{N-1} \left(\begin{matrix}b_k^\dagger & b_{N-k}\end{matrix}\right)\left(\int_0^t\,\mathrm{d}s R_k^\dagger(s) M_k R_k(s)\right) \left(\begin{matrix} b_k\\ b^\dagger_{N-k}\end{matrix}\right).
\label{resultinteraction}
\end{equation}
Performing the integration over $s$ gives
\begin{equation}
\mathcal{O}_J(J,B,t)=\sum_{k=0}^{N-1} \left(\begin{matrix}b_k^\dagger & b_{N-k}\end{matrix}\right)\left(\begin{matrix} \Omega_k & \Delta_k\\ \Delta_k^*  & -\Omega_k\end{matrix}\right) \left(\begin{matrix} b_k\\ b^\dagger_{N-k}\end{matrix}\right),
\label{resultinteraction}
\end{equation}
where
\begin{widetext}
\begin{align}\nonumber
\Omega_k&=\frac{\alpha_k\left(\alpha_k\cos\left(\frac{2\pi k}{N}\right)+\beta_k\sin\left(\frac{2\pi k}{N}\right)\right)t+\frac{\beta_k\left(\beta_k\cos\left(\frac{2\pi k}{N}\right)-\alpha_k\sin\left(\frac{2\pi k}{N}\right)\right)\sin\left(2\omega_kt\right)}{2\omega_k}}{\omega_k^2}\\
\Delta_k&=\frac{i\beta_k\left(\alpha_k\cos\left(\frac{2\pi k}{N}\right)+\beta_k\sin\left(\frac{2\pi k}{N}\right)\right)t+\left(\beta_k\cos\left(\frac{2\pi k}{N}\right)-\alpha_k\sin\left(\frac{2\pi k}{N}\right)\right)\left(\sin^2\left(\omega_kt\right)-\frac{i\alpha_k\sin\left(2\omega_kt\right)}{2\omega_k}\right)}{\omega_k^2}
\label{coefficientsJ}
\end{align}
\end{widetext}
where we have separated the linear and oscillatory behaviour of $\Omega_k$ and $\Delta_k$ (see also~\cite{Pang:14}) and 
$\omega_k=\sqrt{\alpha_k^2+\beta_k^2}$.  As Eq.~\eqref{resultinteraction} is of the same form as 
Eq.~\eqref{Hamiltonianmatrixform} it can be brought to the diagonal form 
\begin{equation}
\mathcal{O}_J(J,B,t)=2\sum_{k=0}^{N-1}\sqrt{\Omega_k^2+\left|\Delta_k\right|^2}\quad d_k^\dagger d_k
\label{diagonalO1}
\end{equation}  
by a suitable Bogoliubov transformation (see Eq.~\eqref{BT}). Note that the free fermions corresponding to 
$d_k^\dagger, \, d_k$ are different from those of Eq.~\eqref{Isingfreefermion}, and we may choose without loss of 
generality the positive square square root of $\Omega_k^2+\left|\Delta_k\right|^2$, which corresponds to choosing the 
vacuum state for free fermions to be zero.  

The optimal precision in estimating the interaction strength $J$ is now easy to determine.  As the latter is inversely 
proportional to the square root of the variance of $\mathcal{O}_J(J,B,t)$, we simply need to determine the maximum achievable 
variance 
for the operator in Eq.~\eqref{diagonalO1}.  This is achieved by preparing the equally weighted superposition of the 
vacuum state and the state where all $N$ modes are occupied by fermions.  The variance with respect 
to this state is simply given by
\begin{equation}
\Delta^2\mathcal{O}_J(J,B,t)_{\max}=\left(\sum_{k=0}^{N-1}\sqrt{\Omega_k^2+\left|\Delta_k\right|^2}\right)^2.
\label{Jvariance}
\end{equation} 
As the sum includes $N$ summands, the variance of $\mathcal{O}_J(J,B,t)$ scales as $N^2$ up to some constant 
factor that depends solely on the ratio between $J$ and $B$ as we now explain.  

\begin{figure}
\includegraphics[keepaspectratio,width=8cm]{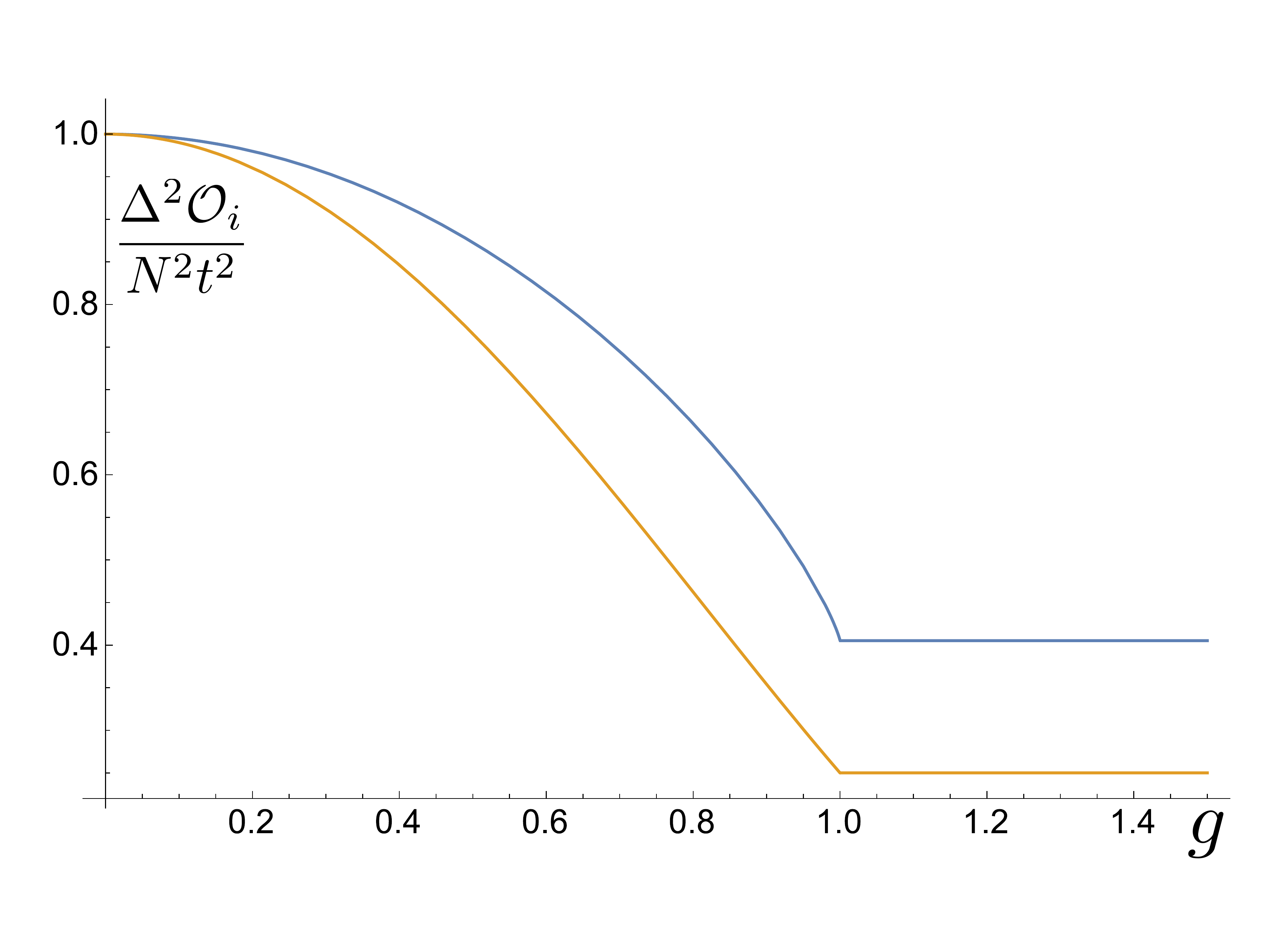}
\caption{The asymptotic value (large $t$ and large $N$) of the normalized variance $\Delta^2\mathcal{O}_i/N^2 t^2$ where 
$i\in (J, B)$, as the function of the parameter ratio $g$. The blue line corresponds to the variance of 
$\Delta^2\mathcal{O}_i/(Nt)^2$ with respect to the optimal state whereas the orange curve corresponds to the constant factor 
$F(g)$ that multiplies the quadratic term of the variance of $\mathcal{O}_i$ with respect to the GHZ state. 
Observe the phase transition at the point where the parameters are equal.}
\label{Asymptotic}
\end{figure}

Using Eq.~\eqref{coefficientsJ} one can easily show that
\begin{widetext}
\begin{equation}
\Omega_k^2+\left|\Delta_k\right|^2=\frac{\left(\beta_k\cos\left(\frac{2\pi k}{N}\right)-\alpha_k\sin\left(\frac{2\pi k}{N}\right)\right)^2\left(1-\cos\left(2\omega_kt\right)\right)+2t^2\omega_k^2\left(\alpha_k\cos\left(\frac{2\pi k}{N}\right)+\beta_k\sin\left(\frac{2\pi k}{N}\right)\right)^2}{2\omega_k^4}
\label{eigenvalues}
\end{equation}
\end{widetext}
For long interaction times, i.e.,~$t\to\infty$ the term quadratic in $t$ in Eq.~\eqref{eigenvalues} completely dominates.  Moreover, for 
$N$ large the sum in Eq.~\eqref{Jvariance} can be replaced, to a good approximation, by an integral resulting in the following simple expression for the variance
\begin{align}\nonumber
&\Delta^2\mathcal{O}_J(J,B,t)_\text{max}= N^2 t^2 \, G\left(\frac{B}{J}\right)\\
&G(g)=\left(\frac{1}{2\pi} \int_0^{2\pi} \sqrt{\frac{(1+ g \cos(x) )^2}{1+ g^2 + 2 g \cos(x)}}dx\right)^2.
\label{Jasymptotic}
\end{align}
Thus, the variance scales as $N^2 t^2$, i.e., at the Heisenberg limit, up to an overall constant factor that only depends 
on the ratio $B/J$. The function $G$ is plotted in Fig.~\ref{Asymptotic}.  Notice that $G(B/J)$ exhibits a phase 
transition at $J=B$. 

\begin{figure}[htb]
\includegraphics[keepaspectratio,width=9cm]{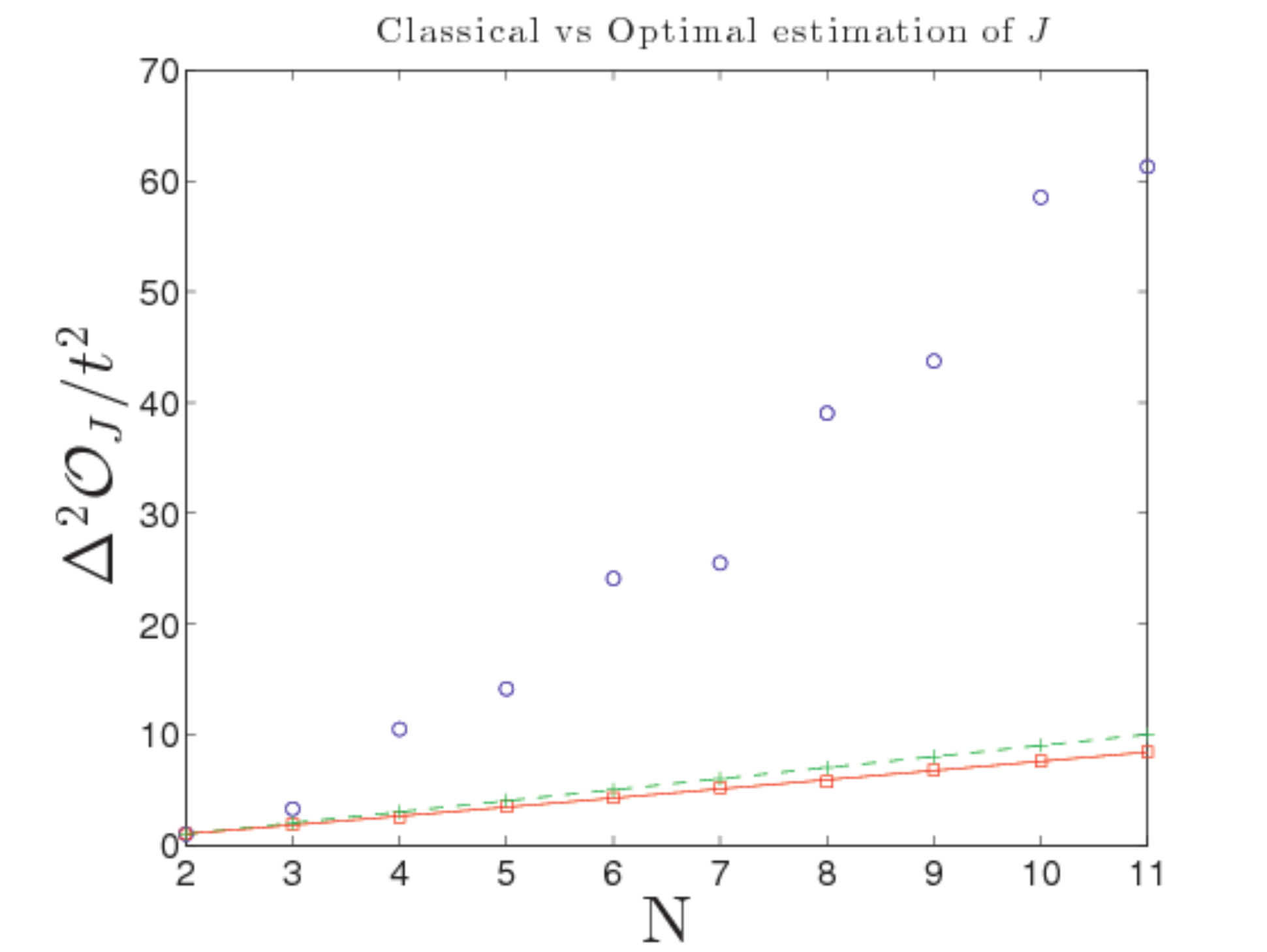}
\caption{Comparison between $\Delta^2\mathcal{O}_J(J,B,t)_{\max}/t^2$ (blue circles) and the variance of 
$\mathcal{O}_J(J,B,t)/t^2$ using product input states (red squares). Both expressions are computed for the case $J=B=1$ and 
$t=20$. The blue circles represent the exact analytical value for $\Delta^2\mathcal{O}_J(J,B,t)_{\max}/t^2$ using 
Eq.~\eqref{coefficientsJ}.  The minimum squared error fit for the red squares is given by $\mathcal{O}_J(J,B,t)/t^2=a N^b+c$ with 
$a=0.7476\pm0.2815,\; b=1.034\pm0.1350,\; c=-0.5139\pm0.5831$ with a $95\%$ confidence. The green line represents the 
optimal QFI for the case where $H=J\sum_{i=1}^{N-1}\sx^{(i)}\sx^{(i+1)}$, i.e., when $B$ in Eq.~\eqref{Ising} is set to zero, using the 
optimal product state $\ket{\psi}=\ket{01}^{\otimes N/2}$.}
\label{productvsoptimalJvariance}
\end{figure}

We now proceed to determine the optimal precision with which one can estimate $J$, knowing $B$, if we restrict the 
input state of $N$ systems to be a product state.  As it is not immediately evident what separable states look like in the 
basis that diagonalizes the operator $\mathcal{O}_J(J,B,t)$, we work directly with Eq.~\eqref{operator}.  
However, due to the form of Eq.~\eqref{operator} it is highly non-trivial to perform an analytical optimization over all 
possible separable states of $N$ qubits.  To that end we perform a brute-force optimization of the variance 
of Eq.~\eqref{Jvariance} over all possible input product states for up to $N=11$ qubits.
The results are shown in Fig.~\ref{productvsoptimalJvariance} where one can already see a difference in scaling between the 
optimal product state strategy and the optimal quantum strategy even for small system sizes.  Moreover, within a high margin of 
certainty the scaling of the QFI using the optimal product input state is at most linear in $N$. 

\subsection{Estimating the field strength}
\label{estimatingB}

We now proceed to estimate the magnetic field $B$, given we know $J$ exactly. The procedure is identical to that of Sec.~\ref{estimatingJ}.   Writing $H_2=\sum_{i=1}^N\sz^{(i)}$ in terms of 
the fermionic operators $b_k$ we obtain
\begin{equation}
H_2=\sum_{k=0}^{N-1}\left(\begin{matrix}b_k^\dagger & b_{N-k}\end{matrix}\right)\left(\begin{matrix}1 &0\\ 0 &-1\end{matrix}\right)\left(\begin{matrix}b_k\\b^\dagger_{N-k}\end{matrix}\right).
\label{h2intermsofb}
\end{equation}
Substituting Eqs.~(\ref{actionHonb},~\ref{h2intermsofb}) into Eq.~\eqref{operator} and 
performing the integration over $s$ yields
\begin{equation}
\mathcal{O}_B(J,B,t)=\sum_{k=0}^{N-1}\left(\begin{matrix}b_k^\dagger & b_{N-k}\end{matrix}\right)\left(\begin{matrix}A_k & B_k\\ B_k^* &-A_k\end{matrix}\right)\left(\begin{matrix}b_k\\b^\dagger_{N-k}\end{matrix}\right),
\label{resultfield}
\end{equation}
where 
\begin{align}\nonumber
A_k&=\frac{\alpha_k^2 t}{\omega_k^2}+\frac{\beta_k^2\sin(2\omega_kt)}{2\omega_k^2}\\
B_k&=\frac{i\alpha_k\beta_k t}{\omega_k^2}+\frac{2\beta_k\omega_k\sin^2(\omega_kt)-i\alpha_k\beta_k\sin(2\omega_kt)}{2\omega_k^2},
\label{coefficientsB}
\end{align}
where we have again separated the linear and oscillatory parts in $t$.  As Eq.~\eqref{resultfield} is of the same form as 
Eq.~\eqref{Hamiltonianmatrixform} it can be brought to the diagonal form
\begin{equation}
\mathcal{O}_B(J,B,t)=2\sum_{k=0}^{N-1}\sqrt{A_k^2+\left|B_k\right|^2}\quad f_k^\dagger f_k
\label{operatorforB}
\end{equation}
by a suitable Bogoliubov transformation (see Eq.~\eqref{BT}).  The optimal achievable precision for estimating the field strength $B$ is given by the optimal variance of 
$\mathcal{O}_B(J,B,t)$ in Eq.~\eqref{operatorforB} which can be easily computed to be 
\begin{equation}
\Delta^2\mathcal{O}_B(J,B,t)_{\max}=\left(\sum_{k=0}^{N-1}\sqrt{A_k^2+\left|B_k\right|^2}\right)^2,
\label{varianceforB}
\end{equation}
and is again achieved by the equally weighted superposition of the vacuum state and the state 
where all $N$ fermionic modes are occupied.  We note that because the Bogoliubov transformation 
diagonalizing Eq.~\eqref{resultfield} explicitly depends on the coefficients of Eq.~\eqref{coefficientsB} the fermions 
described by modes $f_k$ here are different than those described by modes $d_k$ and $c_k$ in 
Eqs.~(\ref{diagonalO1},~\ref{Isingfreefermion}) respectively.  As the sum in Eq.~\eqref{varianceforB} includes $N$ 
summands, the maximum variance of $\mathcal{O}_B(J,B,t)$ scales as $N^2$ up to some factor.

Using Eq.~\eqref{coefficientsB} one can easily show that 
\begin{equation}
A_k^2+|B_k|^2=\frac{\beta_k^2\left(1-\cos\left(2\omega_kt\right)\right)+2t^2\alpha_k^2\omega_k^2}{2\omega_k^2}. 
\end{equation}
For $t\to\infty$ and $N$ very large the quadratic term in $t$ dominates and the maximal variance can be given explicitly as 
\begin{equation}
\Delta^2\mathcal{O}_B(J,B,t)_{\max}=N^2t^2 G\left(J/B\right)
\end{equation}
where $G$ is the function given in Eq.~\eqref{Jasymptotic} and Fig.~\ref{Asymptotic}.   

That the optimal precision in estimating either $B$ or $J$ is asymptotically given by the same expression can be understood 
via the duality of the one-dimensional Ising chain~\cite{Suzuki:13}.  The duality is associated with whether one adopts the spin degrees of freedom 
on the chain or the kink degrees of freedom---associated with the links of the chain---as qubits.  If adjacent spins are 
parallel then there is no kink, else there is a kink.  One can then recast the Ising Hamiltonian of Eq.~\eqref{Ising} in terms of the kink 
degrees of freedom as 
\begin{equation}
H(J,B)= J\sum_{i=1}^N\tau_x^{(i)}+B\sum_{i=1}^{N-1}\tau_z^{(i)}\tau_z^{(i+1)},
\label{Isingkinks}
\end{equation}
where $\{\tau_{\alpha}: \alpha\in(x,y,z)\}$ have the same commutation relations as the Pauli matrices for the spin degree of freedom. 
Up to a basis change, the Hamiltonian in Eq.~\eqref{Isingkinks} is identical to that of Eq.~\eqref{Ising}, except that the 
roles of $J$ and $B$ are reversed.  This is the reason why for the estimation of $J$ the constant factor is given by $G(B/J)$ 
whereas for the estimation of the field it is given by $G(J/B)$.  
We note that the duality works well for spins and kinks in the bulk of the one-dimensional chain but is problematic with spins and 
kinks close to the edges of the chain. However, for large enough $N$ the effects at the boundaries of the chain can be safely 
neglected.  Also note that the quaternionic representation of the operators in the kink degrees of freedom is different than that of 
spins in one crucial way.   The degeneracies of eigenstates in the kink representation are not the same as the ones for spins.  
Indeed, one can easily show that the ordered spectrum of eigenvalues of $\sum_{i=1}^N\sz^{(i)}$ is $\{\lambda_m=\frac{N}{2}-m|\, m
\in(0,\ldots,N)\}$ with eigenvalue $m$ having a degeneracy of $\binom{N}{m}$, whereas the ordered spectrum of 
$\sum_{i=1}^N\tau_x^{(i)}$ is given by $\{\lambda_m=\frac{N}{2}-(m+\frac{1}{2})|\, m\in(0,\ldots, N-1)\}$ with corresponding 
degeneracies given by $2\binom{N-1}{m}$.  Notice however that the duality is not exact for finite $N$.

This difference in the number of degenerate states can be exploited, in the case of 
finite $N$, to improve the precision with which one can estimate parameters in the Bayesian estimation regime.  

\begin{figure}[htb]
\includegraphics[keepaspectratio,width=9cm]{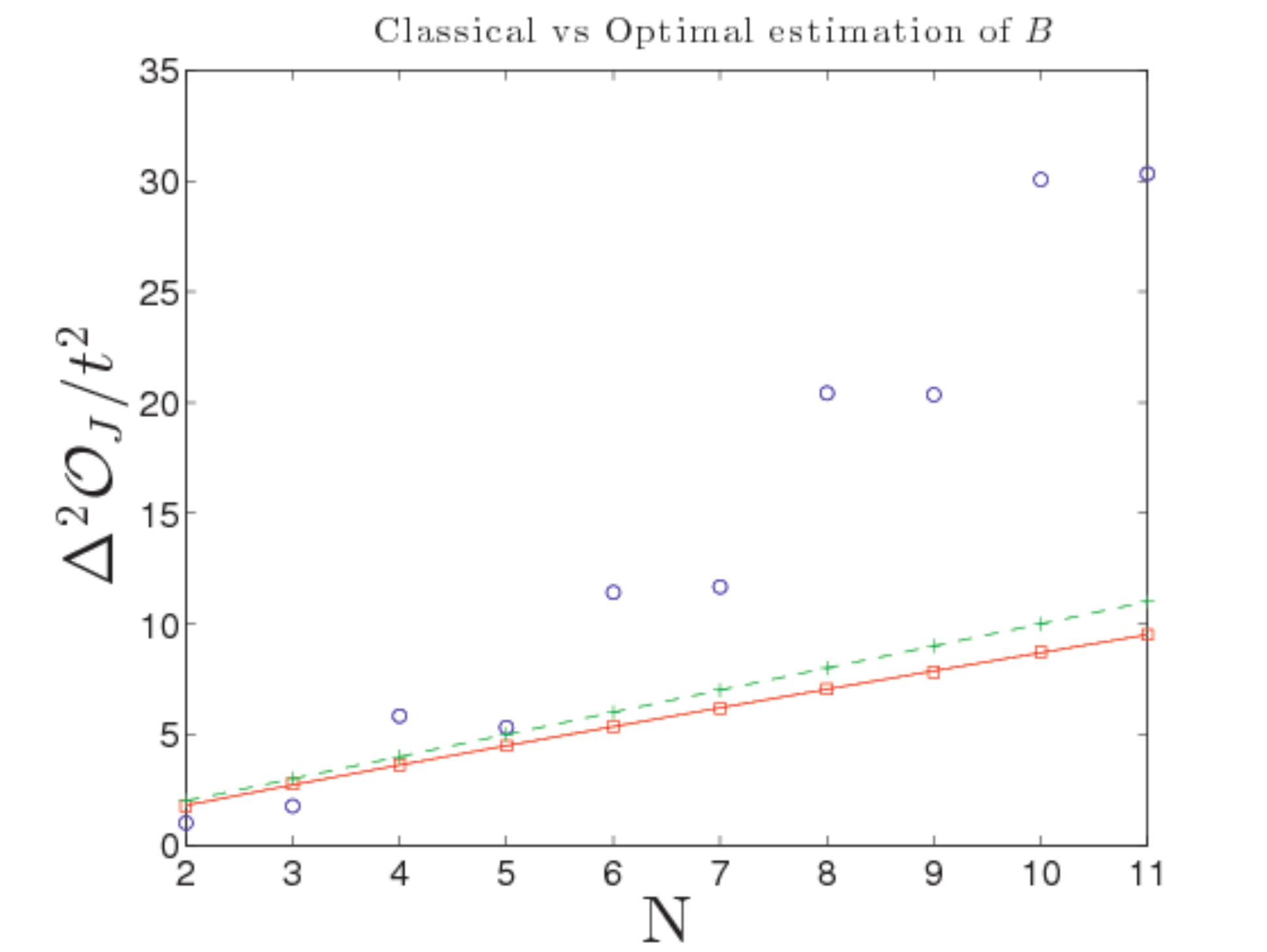}
\caption{Comparison between $\Delta^2\mathcal{O}_B(J,B,t)_{\max}/t^2$ (blue circles) and the optimal variance of 
$\mathcal{O}_B(J,B,t)/t^2$ using product input states (red squares). Both expressions are computed for the case $J=B=1$ and 
$t=20$. The blue circles represent the exact analytical value for $\Delta^2\mathcal{O}_B(J,B,t)_{\max}/t^2$ using 
Eq.~\eqref{coefficientsB}.  The minimum squared error fit for the red squares is given by $\mathcal{O}_B(J,B,t)/t^2=a N^b+c$ with 
$a=1.099\pm0.1350,\; b=0.9114\pm0.0423,\; c=-0.2747\pm0.2421$ with a $95\%$ confidence. The green line represents the 
optimal QFI for the case where $H=B\sum_{i=1}^{N}\sz^{(i)}$, i.e., when $J$ in Eq.~\eqref{Ising} is set to zero, using the optimal 
product state $\ket{\psi}=\ket{+}^{\otimes N}$.}
\label{productvsoptimalvarianceB}
\end{figure}

The optimal precision for estimating $B$, given we know $J$, using a separable strategy is again numerically 
calculated for up to $N=11$ qubits with the results shown in Fig.~\ref{productvsoptimalvarianceB}.
Just as in the case of estimating $J$, one can already observe a difference in scaling of the QFI between the optimal 
product state strategy and the corresponding optimal quantum strategy.  Moreover, within a high margin of certainty, 
the scaling of the QFI using the optimal product state is at most linear in $N$.

\section{Precision using GHZ type states}
\label{GHZprecision}

In the previous section we showed that the optimal precision in estimating either the interaction strength or magnetic field is 
achieved by states that are linear superpositions of the vacuum and all $N$ sites occupied by fermions. Whereas such 
states can be prepared efficiently, i.e., with a quantum circuit that grown polynomially with the number of qubits~\cite{Verstraete:09}, preparing such states in practice may still be quite challenging due to the optimal states dependence 
on both time, and $J$ ($B$).  An important question, then, is whether there exist states that are easy to prepare and yield 
Heisenberg scaling in precision for all time and all values of $J$ ($B$). For example in ion-trap set-ups one can prepare 
the GHZ state using a 
single S\o rensen-M\o lmer gate~\cite{Monz:11}.  In this section we analytically
determine the performance of GHZ-type states for estimating either the interaction strength or magnetic field. 
We will show that the GHZ-type states allow for Heisenberg scaling in precision for all values of $J$ ($B$), with 
only a constant factor difference from the optimal precision achievable.

Let us first determine the ultimate precision of estimating $B$ using the GHZ state
\begin{equation}
\ket{GHZ}=\frac{1}{\sqrt{2}}\left(\ket{0}^{\otimes N}+\ket{1}^{\otimes N}\right).
\label{ghz}
\end{equation}
The GHZ state yields the ultimate precision in estimating $B$ for the case where the latter is imprinted via the unitary operator 
$U=\exp(-i B t H_2)$.  In order to calculate its performance for the Ising Hamiltonian of Eq.~\eqref{Ising} we will use the Jordan-
Wigner transformation and the discrete Fourier transform to express the GHZ state in terms of the fermionic creation operators $b_k^\dagger$.   

We begin by noting that the GHZ state can be written in terms of the fermionic operators $\{a,\,a^\dagger\}$ as
\begin{equation}
\ket{GHZ}=\frac{1}{\sqrt{2}}\left(\one+\prod_{k=1}^Na_k^\dagger\right)\ket{\bm{0}}.
\label{JWGHZ}
\end{equation}  
Using Eq.~\eqref{FT} to transform the fermionic operators, $\{a_k^\dagger\}$ to $\{b_k^\dagger\}$, and noting that the Fourier 
transform leaves the vacuum invariant, yields
\begin{equation}
\ket{GHZ}=\frac{1}{\sqrt{2}}\left(\ket{\bm{0}}+e^{i\phi}\ket{\bm 1}\right),
\label{FTGHZ1}
\end{equation}
where $\phi$ is a relative phase that arises due to the Fourier transform.

Using Eq.~\eqref{FTGHZ1} to calculate the variance of the operator in Eq.~\eqref{resultfield}, one finds that 
\begin{align}\nonumber
\bra{GHZ}\mathcal{O}_B(J,B,t)\ket{GHZ}&=0\\
\bra{GHZ}\mathcal{O}^2_B(J,B,t)\ket{GHZ}&=\left(\sum_{k=0}^{N-1} A_k\right)^2,
\label{expectations}
\end{align}
and consequently $\Delta^2\mathcal{O}_B(J,B,t)=\bra{GHZ}\mathcal{O}^2_B(J,B,t)\ket{GHZ}$.  Using 
Eq.~\eqref{coefficientsB}, 
denoting $\theta=\frac{2\pi k}{N}$, and letting $N\to\infty$ the variance of $\Delta^2\mathcal{O}_B(J,B,t)$ explicitly reads
\begin{equation}
\Delta^2\mathcal{O}_B(J,B,t)=\frac{N^2}{4\pi^2}\left(\int_0^{2\pi}\frac{(1+g\cos\theta)^2}{1+g^2+2g\cos\theta}\mathrm{d}\theta\right)^2
\label{ghzvariance1}
\end{equation}
where $g=J/B$.  From Eq.~\eqref{ghzvariance1} it follows that the variance of $\mathcal{O}_B(J,B,t)$ with respect to the 
GHZ state always scales quadratically with $N$, i.e., at the Heisenberg limit, with a prefactor $F(g)$ slightly lower than that 
for the optimal states as shown in Fig.~\ref{Asymptotic}.  

We now determine the precision with which one can estimate $J$  using the state
\begin{equation}
\ket{\psi}=\left(\frac{\ket{00}^{\otimes{\frac{N}{2}}}+\ket{01}^{\otimes{\frac{N}{2}}}}{\sqrt{2}}\right),
\label{staggeredGHZ}
\end{equation}
which is, up to a basis change,  the state that yields the ultimate precision in estimating $J$ for the case where the latter is imprinted via the unitary dynamics $U=\exp(-iJH_1)$.  We 
note that since $H_1$ has a doubly degenerate eigenspace for both its minimum and maximum eigenvalue, the optimal state for 
estimating $J$ is not unique.  

If we express this state in terms of the kink 
degrees of freedom, as opposed to the spin degrees of freedom then what we obtain is, up to an overall Hadamard transformation, 
the GHZ state, i.e, a linear superposition of zero kinks, and $N$ kinks.  As the Ising Hamiltonian expressed in the kink degrees of 
freedom is given by Eq.~\eqref{Isingkinks}, up to an overall Hadamard transformation, it follows that the variance of 
$\mathcal{O}_J(J,B,t)$ with respect to the state $\ket{\psi}$ is given by Eq.~\eqref{ghzvariance1}, with $g=B/J$.

\section{Conclusion}
\label{Conclusion}

In this work we investigated precision limits for noiseless quantum metrology in the presence of parameter dependent 
Hamiltonians, and in particular the Ising Hamiltonian.  We showed that the ultimate limit in estimating the interaction 
or magnetic field strength scales quadratically with the number of probe systems $N$, i.e., at the Heisenberg limit, 
and that the states that achieve this precision are linear superposition of the vacuum and fully occupied states 
of $N$ free fermions. Moreover, whereas the Ising Hamiltonian generates entanglement this entanglement does not 
help in boosting the precision scaling with respect to $N$ that can 
be achieved with product states.  In addition, we showed that the achievable precision in estimating either $J$ or $B$ for the Ising 
Hamiltonian using GHZ-type states also scales at the Heisenberg limit. 

Whereas we have shown that the entanglement generating properties of the Ising Hamiltonian do not boost the 
precision in estimation for product states, it may still be the case that we can exploit this property of the Ising 
Hamiltonian to reduce the amount of entanglement required in the initial input state of the $N$ probes.  This would be of high 
interest for practical realizations of quantum metrology where the creation of highly entangled states of $N$ systems remains a 
challenge.

In addition, our analysis deals with optimal states and bounds in the absence of noise.  It would be interesting to 
investigate the achievable precision bounds in the presence of several physical noise models, such as uncorrelated 
dephasing or depolarizing noise, as well as spatial and temporal correlated noise.  Furthermore, it would be interesting to determine 
which of these types of noise can we readily combat via the use of error-correcting techniques, or by dynamical 
decoupling~\cite{Dur:14, Arad:14, Kessler:14}.   

\section{Acknowledgements}
The authors would like to thank the anonymous referee for his valuable comments and suggestions during the reviewing 
process of this article. This work was supported by the Austrian Science Fund (FWF): P24273-N16 and the Swiss National Science Foundation grant P2GEP2\_151964.
  
\bibliographystyle{apsrev4-1}
\bibliography{Ising}

\end{document}